\documentclass[twocolumn,prb,aps,showpacs,preprintnumbers,amsmath,amssymb]{revtex4}

\usepackage{graphicx}
\usepackage{dcolumn}
\usepackage{bm}
\usepackage{epsf}

\newcommand{\beq}{\begin{equation}}
\newcommand{\eeq}{\end{equation}}
\newcommand{\beqa}{\begin{eqnarray}}
\newcommand{\eeqa}{\end{eqnarray}}
\newcommand{\beqan}{\begin{eqnarray*}}
\newcommand{\eeqan}{\end{eqnarray*}}
\newcommand{\ben}{\begin{enumerate}}
\newcommand{\een}{\end{enumerate}}
\newcommand{\bit}{\begin{itemize}}
\newcommand{\eit}{\end{itemize}}

\begin{document}


\title{A numerical investigation of a piezoelectric surface acoustic wave interaction with a one-dimensional channel\\}

\author{S. Rahman$^1$}
\email{S.Rahman.00@cantab.net}

\author{M. Kataoka$^1$}%
\author{C. H. W. Barnes$^1$}%
\author{H. P. Langtangen$^2$}%
\affiliation{$^{1}$Cavendish Laboratory, Madingley Road, Cambridge, CB3 0HE, United Kingdom}%
\affiliation{$^2$Simula Research Laboratory, Martin Linges v 17,
Fornebu P.O.Box 134, 1325 Lysaker, Norway}%

\date{\today}

\begin{abstract}
We investigate the propagation of a piezoelectric surface acoustic
wave (SAW) across a GaAs/Al$_X$Ga$_{1-X}$As heterostructure
surface, on which there is fixed a metallic split-gate. Our method
is based on a finite element formulation of the underlying
equations of motion, and is performed in three-dimensions fully
incorporating the geometry and material composition of the
substrate and gates. We demonstrate attenuation  of the SAW
amplitude as a result of the presence of both mechanical and
electrical gates on the surface. We show that the incorporation of
a simple model for the screening by the two-dimensional electron
gas (2DEG), results in a total electric potential modulation that
suggests a mechanism for the capture and release of electrons by
the SAW. Our simulations suggest the absence of any significant
turbulence in the SAW motion which could hamper the operation of
SAW based quantum devices of a more complex geometry.

\end{abstract}

\pacs{85.35.Gv, 73.23.-b, 77.65.Dq}
\maketitle

\section{\label{sec:intro}INTRODUCTION}

Surface acoustic waves (SAWs) are widely used in microwave circuit
components, such as filters and resonators.\cite{Matthews} In
condensed matter physics research, SAWs have been a useful tool in
probing electronic structure, for example, in thin metal
films,\cite{Bierbaum:72} and quantum Hall
liquids,\cite{Wixforth:86} for a number of decades.  In 1996,
Shilton \emph{et al.} realized a device which carries a quantized
number of electrons through a quasi-one-dimensional channel (Q1DC)
using SAWs.\cite{Shilton:96} Such devices are currently being
developed for metrological applications and for quantum logic
circuits.\cite{BarnesQC1}

Despite the enthusiasm for using SAWs in such applications,
experimental progress in this direction has been slow. One of the
main reasons for this is that we lack a rigorous understanding of
the dynamics of the SAW propagation in complicated device
structures, and are therefore unable to precisely design or
simulate the devices. Earlier analytical work on SAW
single-electron transport (SET)
devices\cite{Aizin:98,Gumbs:98,Gumbs:99} required crude
approximations, especially regarding the effect of surface gates,
where a realistic two-dimensional model of a split-gate device had
not been achieved.

In this paper, we present the results of a numerical investigation
of the dynamics at the depth of a two-dimensional electron gas
(2DEG) in a GaAs/Al$_{X}$Ga$_{1-X}$As heterostructure, resulting
from the propagation of a piezoelectric SAW through a realistic
split-gate. The numerical method implemented,\cite{Rahman:05}
fully takes into account the geometry and material composition of
the device, as well as the full two-way coupling between the
electrical and mechanical fields. Our simulations are performed in
the strongly screened, low SAW power, low barrier height, regime,
as this allows us to implement a simple model for the 2DEG and
allows assumptions on the SAW amplitude and split-gate barrier
height.

We use our solutions to discuss the effects due to the presence of
mechanical and electrical surface gates. Although in this paper we
restrict our attention to a split-gate device, our method is in
principle applicable to a device of any geometry.

This paper proceeds as follows. In Section~\ref{sec:theory}, we
describe the details of the underlying theory of SAW propagation
in a gated device.  We discuss the earlier work of Aizin and Gumbs
\cite{Aizin:98,Gumbs:98} pointing out the differences between our
formalisms and solutions. In Section~\ref{sec:method}, we describe
the method behind our numerical strategy.  The results for an
un-gated device and a split-gate device are presented in
Sections~\ref{sec:resultsa} and \ref{sec:resultsb}, respectively.

\section{\label{sec:theory}SINGLE-ELECTRON SAW DEVICES}

\begin{figure}

\epsfxsize=8.5cm

\centerline{\epsffile{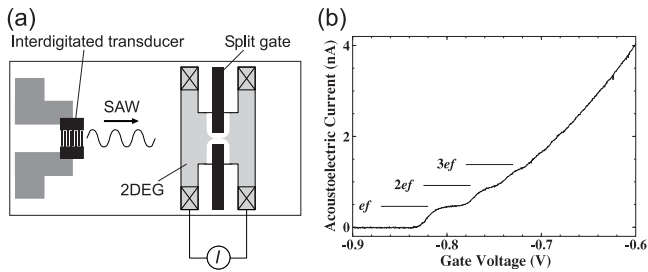}}

\caption{(a) A schematic diagram showing the experimental setup in
acoustic charge transport experiments. A transducer on the left
excites a SAW wave that propagates towards the spit-gate on the
right. (b) The acoustoelectric current versus split-gate voltage.
The current exhibits a step-like behavior as the gate voltage is
varied. } \label{fig:SAWDevice}
\end{figure}

The quantization of acoustoelectric current was first demonstrated
experimentally by Shilton \emph{et al}.
\cite{Shilton:96,Shilton:96a} Figure~\ref{fig:SAWDevice}(a) shows
the experimental setup used. It consists of a
GaAs/Al$_{X}$Ga$_{1-X}$As heterostructure containing a
two-dimensional electron gas (2DEG) that is formed into a mesa by
wet etching. A Q1DC is formed when a negative potential is applied
to a split metallic surface gate. A large negative voltage applied
to the gates induces a narrow depleted constriction between the
two 2DEG regions. Application of a microwave signal to an
interdigitated transducer excites electromechanical waves through
the piezoelectric effect which include both SAWs and bulk waves.

The electric component of the SAW captures electrons from the 2DEG
in its potential minima transporting them through the constriction
produced by the split-gate potential. The current measured from
the drain region exhibits a step-like behavior as shown in
Fig.~\ref{fig:SAWDevice}(b). The values of the current plateaus
are quantised with \beq I = nef, \eeq \noindent where $I$ is the
current, $e$ is the electron charge, $n$ is the number of
electrons transported, and $f$ is the SAW frequency. The case with
$n=1$ involves the transport of a single electron per cycle
through the Q1DC.

The original explanation for the quantization of acoustoelectric
current given by Shilton \emph{et
al.}\cite{Shilton:96,Shilton:96a} asserts that the combination of
the SAW electrostatic potential and the split-gate potential
produces a travelling quantum dot that transports a fixed number
of electrons from one side of the constriction to the other. This
is plausible as the length of the depleted region formed by the
split-gate ($\thicksim$ $1.5~\mu$m) is a small fraction greater
than SAW wavelength ($\thicksim$ 1~$\mu$m). Several electrons are
believed to be initially trapped in each SAW minimum but as it
passes through the constriction formed by the split-gate, its
physical dimensions decrease and Coulomb repulsion between
electrons forces electrons out, thus determining the final number
of electrons remaining in the dot. This explanation, although
qualitatively satisfactory, was not at the time, supported by a
detailed theoretical analysis involving an accurate model for the
SAW and split-gate, which we aim to develop in this paper.

One approach for such an analysis, is to solve the equations of
motion in the GaAs/Al$_X$Ga$_{1-X}$As media, taking into account
the complex geometry of the split-gate, the material composition
and external electric fields, to determine the evolution of the
SAW potential through the device. The resulting electric potential
could be used as a part of a quantum-mechanical treatment to
describe the evolution of the electronic states.

However, this approach is a non-trivial one, as the equations of
motion in a piezoelectric material such as Al$_X$Ga$_{1-X}$As,
consist of a set of four coupled partial differential equations,
second order in time and space. Moreover, as the SAW passes
through the split-gate potential, initially $\sim 30$ electrons
are captured from the 2DEG, this number reduces as the SAW quantum
dot is dragged into the increasing electric field in the
constriction. This problem then involves a solution to the
time-dependent many-particle Schr\"{o}dinger equation including
spin. To perform this calculation exactly is beyond the scope or
intention of this paper. The focus of this paper is the solution
of the equations of motion in the GaAs/Al$_{X}$Ga$_{1-X}$As medium
and, using a simple model for the 2DEG and, ignoring any self
consistent effects, the calculation of the SAW potential as it
traverses the split-gate. The equations of motion in a
heterogeneous piezoelectric material are,\cite{Matthews, Auld}

\begin{equation}\label{eqn:elastMotion}
     \varrho \frac{\partial^2u_i}{\partial t^2} =  \frac{\partial}{\partial x_j} c_{ijkl}^E \frac{\partial u_l}{ \partial x_k} +
    \frac{\partial }{\partial x_j} e_{kij}  \frac{\partial \phi}{\partial x_k},
\end{equation}

\begin{eqnarray}\label{eqn:PiezoPoisson}
 \frac{\partial }{\partial x_i} \epsilon^S_{ij} \frac{\partial \phi}{\partial x_j}
=  \frac{\partial }{\partial x_i} e_{ijk} \frac{\partial
u_k}{\partial x_j},
\end{eqnarray}

\noindent  where $u_i$ is a three-component vector representing
the displacement of the material at each point, $\phi$ is the
electric potential, $c^E_{ijkl}$ are components of the elastic
tensor, $\epsilon^S_{ij}$ ($=\epsilon^S$) are components of the
material dielectric (the superscripts $E$ and $S$ indicate that
the values were determined under constant electric field and
strain, respectively), $\varrho$ is the mass density and $e_{ijk}$
are components of the piezoelectric tensor representing the
coupling strength between the electric and mechanical fields. The
Eqs.~(\ref{eqn:elastMotion}) can be regarded as a system of wave
equations with a load term due to the electric field, and
Eq.~(\ref{eqn:PiezoPoisson}) can be regarded as a Poisson equation
with a source term due to the mechanical deformations.

The equations are solved with traction-free boundary conditions.
For example  $\sigma_{iz} = 0$ for the surface normal to the $z$
axis, where $\sigma_{ij}$ are components of the stress tensor,
defined by

\begin{equation}\label{eqn:Tij1}
    \sigma_{ij} = -e_{kij} E_k + c^E_{ijkl} \varepsilon_{kl},
\end{equation}

\noindent where $E_k = -\frac{\partial \phi}{\partial x_k}$ and
$\varepsilon_{ij}  = \frac{1}{2} (\frac{\partial u_i}{\partial
x_j} + \frac{\partial u_j}{\partial x_i})$. The electrostatic
boundary conditions require the prescription of the normal
component of the electric displacement at the free surface of the
medium. The electric displacement vector is defined by

\begin{equation}\label{eqb:eBC1}
    D_i = \epsilon_{ij}^S E_j + e_{ijk} \varepsilon_{jk}.
\end{equation}

Aizin \emph{et al.}\cite{Aizin:98} followed this approach
providing a closed form \emph{analytic} formula for the combined
potential of the SAW and split-gate. The calculation involved the
simplification of Eqs.~(\ref{eqn:elastMotion}) and
(\ref{eqn:PiezoPoisson}), and of the description of the physical
system, so that analytical solutions were tractable. In
particular, their approach was two-dimensional, taking into
account the direction of SAW propagation and the direction into
the bulk of the heterostructure only. Clearly, this analysis
cannot take into account the geometry of the surface gates and
therefore the `split' nature of the split-gate. In fact, a
split-gate was approximated to be a single bar gate. Another
simplification was the decoupling of the mechanical motion from
the electric fields by setting the piezoelectric term in
Eq.~(\ref{eqn:elastMotion}) to zero. This simplification was based
on the fact that $e_{ijk} \ll c_{ijkl}$ for Al$_X$Ga$_{1-X}$As. In
practice, however, externally applied voltages (and hence electric
fields) may be large enough that they contribute to the mechanical
strains and hence the observed potential in the vicinity of the
gates.\cite{Rahman:05} Thus, the term with $e_{ijk}$ in
Eq.~(\ref{eqn:elastMotion}) cannot always be neglected. Their
calculations also ignore differences in the elastic constants and
mass density between the gates and the heterostructure substrate.
The gates are often fabricated from a combination of aluminium
(Al), gold (Ag) or titanium (Ti). In fact, this approach neglects
the mechanical presence of the gates at all: the SAW potential
$\phi_{SAW}$ is effectively assumed to have a cosine form i.e.
$\phi_{SAW} \sim \cos(k x - \omega t)$ and is  added to the
split-gate potential $\phi_{SG}$ which is of a quadratic form, to
give the total electric potential $\phi$. In practice however, the
mechanical presence of surface gates could cause the SAW potential
to be damped, reflected or diffracted in the vicinity of the
gates. The presence of surface gate screening has been utilized in
SAW-based photoluminescence experiments.\cite{Alsini:03,Sogawa:01}

However, using the potential $\phi$, Aizin and Gumbs were able to
show that if tunnelling of electrons from the quantum dot to the
source 2DEG was allowed, they could explain the transition in the
acoustoelectric current from the point where no charge is
transported, to that where a single electron is
transported.\cite{Aizin:98} In effect, they provided an
explanation of the first current plateau. In subsequent
theoretical work,\cite{Gumbs:99} Gumbs \emph{et al.} showed, using
the same potential, that the second quantized plateau in the
current as a function of the gate voltage or SAW power can be
explained by the effect of both the Coulomb blockade in the
quantum dot and the backward tunnelling into the Q1DC.

The analytic approach by Aizin and Gumbs was a convenient one, and
similar approaches for the SAW and gate potential have been
adopted in other theoretical
works.\cite{RobinsonBarnes,Maksym:00,Flensberg:99}

Our recent numerical investigation,\cite{Rahman:05} based on the
finite element formulation of Eqs.~(\ref{eqn:elastMotion}) and
(\ref{eqn:PiezoPoisson}) demonstrated both the effect of the
mechanical presence of a gate on the SAW electric potential and
the presence of a large static electric potential on the
mechanical strains in the material. The procedure includes the
full two-way coupling between the electric and mechanical fields.
The advantage of the approach is that we can handle the geometry
and material composition of the split-gate, or any shape of gate
pattern, easily. In particular, the SAW wavefront was shown to be
damped and scattered somewhat, after propagating through a single
gate. Such effects could prevent the proper functioning of SAW
devices, operating in the quantum regime. However, the intention
of that paper was to demonstrate that the numerical method is
capable of reproducing the fully coupled elasto-electric dynamics
of the physical system. In order to exaggerate the effects of the
electrical and mechanical coupling, the investigation used
fictitious materials for the gate, which had substantially
different physical properties to Al, Ti or Ag, which are
frequently used in SAW based SET experiments.

In this paper, we apply our numerical solution strategy to a
split-gate device with realistic parameters for the gate geometry
and material composition. We also include a simple model for the
2DEG, when a voltage is applied to the split-gate.

\section{\label{sec:method}METHOD}

\begin{figure}

\epsfxsize=7.0cm

\centerline{\epsffile{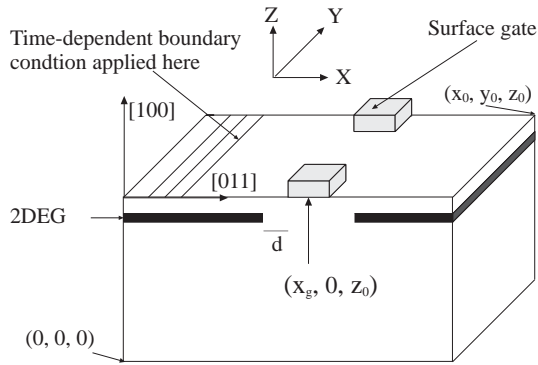}}

\caption{A schematic diagram showing the device geometry. A
split-gate, composed of a Ti/Al alloy, is placed on the surface
($z=z0$) of the GaAs/Al$_{X}$Ga$_{1-X}$As substrate. The SAW,
generated by the three membranes separated 1.0~$\mu$m apart,
propagates along the $x$-axis. The $x_0$, $y_0$ and $z_0$
parameters are chosen to be $12000$~nm, $3000$~nm and $8000$~nm
respectively. The depletion parameter $d$ is $300$~nm. The gate is
centered along the $x$ axis at $x = x_g = 9850$~nm.}
\label{fig:model}

\end{figure}

Figure~\ref{fig:model} illustrates the geometry of the device to
be modelled. The device consists of a GaAs/Al$_{X}$Ga$_{1-X}$As
($X=0.3$) heterostructure with a Ti/Al split-gate placed on the
surface. The sample has dimensions $12$~$\mu$m $\times$ $3$~$\mu$m
$\times$ $8$~$\mu$m in the $x$, $y$ and $z$ directions,
respectively, with $z=z_0$ ($z_0$ = $8$~$\mu$m) at the top surface
and the heterostructure occupying the region shown in
Fig.~\ref{fig:model}. From the top surface down, there is a
$10$~nm GaAs capped layer, an $80$~nm Al$_{X}$Ga$_{1-X}$As layer,
and the remainder being GaAs. We are interested in calculating the
dynamic effects of the SAW propagating through the constriction
and therefore we ignore static charge distributions due to doping
in the Al$_{X}$Ga$_{1-X}$As layer. The thickness (or height) of
the Ti and Al components of the gates are chosen to be $20$~nm and
$40$~nm, respectively. The constriction formed by the split-gate
has dimensions 0.7~$\mu$m and 1.0~$\mu$m in the $x$ and $y$
directions, respectively. The gate is centered along the $x$ axis
at $x = x_g = 9850$~nm.

To obtain an accurate description of the potential landscape in
the device (without a SAW) and in particular at the 2DEG, a
three-dimensional Hartree or density functional calculation would
be necessary,\cite{Laux:88} incorporating suitable boundary
conditions for the split-gate and charge from donor impurities.
Moreover, to incorporate the time-varying polarization charge
induced from the SAW, this calculation would need to be performed
at every time-step. Such a task would require enormous
computational resources. Instead, a simple model for the
split-gate induced potential landscape is implemented here, based
on our experimental observations. In the experiments, the
split-gate potentials are applied relative to the 2DEG. The 2DEG
is believed to be depleted a few hundred nano-meters laterally
away from the split-gate. Hence, in this model, when a gate
voltage is applied, the 2DEG is assumed to be depleted $300$~nm
laterally from the split-gate, and is modelled as a metal sheet,
with a potential difference applied between the split-gate and the
metal sheet (with the potential at the 2DEG set to 0~V), in the
numerical solution of the Eq.~(\ref{eqn:elastMotion}). For the
electrical boundary conditions at the free surface, we choose a
Neumann-type condition on the un-gated regions, implemented by
setting $D_z = 0$ in order that overall charge neutrality in the
device is satisfied.\cite{Davies:95,ft2} Dirichlet-type conditions
are unsuitable for the un-gated regions as they would compromise
the SAW potential. The mechanical boundary conditions implemented
are the `traction-free' conditions defined earlier.

 \begin{table*}[t]\label{tab:data}
 \centering
 \begin{tabular}{|c|c|c|c|c|c|c|}
 \hline
 & $c_{11}$ & $c_{12}$ & $c_{44}$ & $e_{14}$ & $\epsilon$ & $\rho$ \\
 Material & ($10^{10}$Nm$^{-2}$)& ($10^{10}$Nm$^{-2}$)& ($10^{10}$Nm$^{-2}$)& (Cm$^{-2}$) & (Fm$^{-1}$) & ($10^3$ Kgm $^{-3}$) \\
 \hline
 Al$_{X}$Ga$_{1-X}$As   & $11.88 + 0.14X$ & $5.38 + 0.32X$    & $5.94 - 0.05X$ & -0.16-0.065X      &  (13.18-3.12X) $\epsilon_0$    &  5.36-1.6X   \\
 Titanium               & 20.30           & 11.47             & 4.416          & n/a   & n/a &  4.540             \\
 Aluminium              & 11.09           & 5.842             & 2.626          & n/a   & n/a &  2.698        \\
 \hline
 \end{tabular}
 \caption{Physical properties of Al$_X$Ga$_{1-X}$As, \cite{Adachi}
 Al \cite{Smithells} and Ti\cite{Smithells}. The permittivity of vacuum is $\epsilon_0$.}
 \end{table*}

In order to excite the SAWs, we apply a time-dependent boundary
condition of the form

\begin{eqnarray}\label{eqn:DispBC}
    u_z &=& A \sin (2 \pi f t),  \qquad x=const, \qquad z=z_0,
\end{eqnarray}

\noindent where $f$ is the frequency and $A$ is the amplitude.
Here, $f$ is chosen to be 2.7~GHz in order to satisfy the relation
$v_{SAW} = f \lambda$, where $v_{SAW}$ is the SAW velocity ($\sim
2.7 \times 10^3$~ms$^{-1}$) and $\lambda$ is the period of the
transducer fingers. The amplitude $A$ is chosen such that the SAW
has the amplitude of $\sim 20$~mV at a depth of 100~nm, and
corresponds to the low SAW power regime of the experiments. The
oscillation is applied at three values of $x$ separated by
$1.0~\mu$m as shown in Fig.~\ref{fig:model}, to increase the SAW
amplitude in the $x$ direction relative to $z$ direction: the SAW
wave-fronts generated by each membrane add more constructively
along the $x$-axis than the $z$-axis.

For Al$_{X}$Ga$_{1-X}$As, the only non-zero components of the
piezoelectric tensor have the value $e_{14}$. Also, the
non-vanishing components of the permittivity tensor are
$\epsilon_{11} = \epsilon_{22} = \epsilon_{33} = \epsilon_{s}$,
and the non-vanishing components of the elastic tensor (not
written out above) are $c_{xxxx}$ = $c_{yyyy}$ = $c_{zzzz}$ =
$c_{11}$, $c_{xxyy}$ = $c_{yyzz}$  = $c_{zzxx}$ = $c_{12}$, and
$c_{xyxy}$ = $c_{yzyz}$ = $c_{zxzx}$ = $c_{44}$. All other
non-zero components of the elastic tensor can be determined by
applying its symmetry properties $c_{ijkl} = c_{jikl} = c_{ijlk} =
c_{klij}$. The values of these constants for the relevant
materials are given in Table~I. The dielectric constant and
$e_{14}$ are not required for the Ti and Al as they are subject to
Dirichlet boundary conditions on all external surfaces. In our
simulations, we follow the convention in SAW based SET experiments
where the crystal orientation is such that the $x$ direction is
aligned to [011] direction, and $z$ direction to [100]
direction.\cite{rotation}

The finite element method was chosen for its proven ability in
handling geometrically complicated
domains.\cite{Zienkiewicz,DpBook2} The basic idea of the finite
element method is to approximate the unknown fields, for example
$\phi$ in the Poisson equation above, by a linear combination of
basis functions $N_i$, $\phi \approx \hat\phi = \sum_{j=1}^n
N_j\phi_j$, then insert ${\hat\phi}$ in the Poisson equation, and
demand the residual to be orthogonal to the space spanned by $\{
N_1,\ldots,N_n\}$. We utilize finite elements in the spatial part
of the problem and use a second order finite difference
discretisation in time for computational speed so
Eq.~(\ref{eqn:elastMotion}) is approximated by

\beq \varrho { u^{\ell -1} -2u^\ell +u^{\ell+1}\over\Delta
t^2} = \text{RHS}^\ell, \label{tEL:eq1} \eeq

\noindent where $\text{RHS}^l$ is a finite element approximation
of the elastic stress term and the mechanical field load. The
superscript $\ell$ represents the time level. We use eight-noded
brick elements corresponding to linear basis functions $N_i$,
resulting in an overall spatial and temporal convergence rate for
the error, of 2.0. An operator splitting strategy is employed to
split the coupled $u_i$-$\phi$ problem.\cite{DpBook2} Inheritance
and polymorphism principles from object-oriented
programming,\cite{DpBook2} are used to incorporate software
components from the Diffpack library, for solving the Poisson and
the elasticity equations, respectively, thus maximizing efficiency
in the programming and verification. The numerical formulation and
verification is described in more detail in
Ref.~\onlinecite{Rahman:05}

\section{\label{sec:results}RESULTS AND DISCUSSION}

\subsection{\label{sec:resultsa}SAW through an un-gated surface}

\begin{figure}

\epsfxsize=8.5cm

\centerline{\epsffile{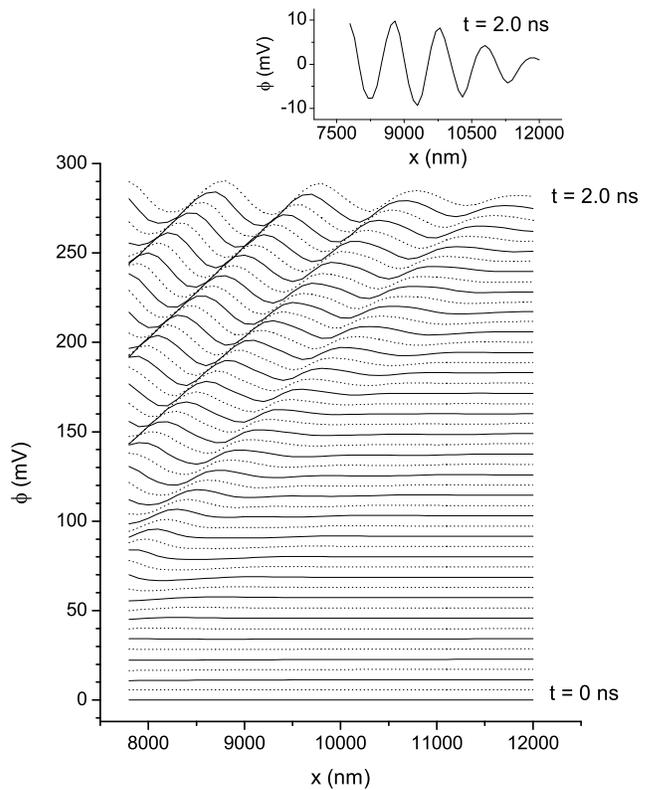}}

\caption{Time development of SAW induced electric potential $\phi$
as a function of the $x$-axis (without gates on the surface and
without the 2DEG). For clarity, the inset shows the amplitude of
the SAW is $\sim 20$~mV.}

\label{fig:potential_X_series}

\end{figure}

We first demonstrate that our method for exciting acoustoelectric
vibrations using three vertically oscillating membranes, excites
the SAW modes with the required wavelengths and velocity, by
performing a simulation without any surface gates and without the
2DEG. Therefore, it is not necessary to set the total electric
potential $\phi$ to $0$~V at the 2DEG in this simulation. (Also,
the absence of depletion would result in the SAW being invisible
everywhere at that depth.) Figure~\ref{fig:potential_X_series}
shows the time development of the electric potential as a function
of distance along the propagation direction ($+x$ direction). The
curves were extracted at a depth of $100$~nm (where the
interesting physics of acoustoelectric charge transport take
place). The time-dependent boundary condition given by
Eq.~(\ref{eqn:DispBC}) is applied on the left-hand-side of
Fig.~\ref{fig:potential_X_series}, with the closest membrane at
$x=6600$~nm (not shown). We see that two transient peaks, clearly
distinguishable from all the other peaks, are initially formed,
and are followed by a more consistent set of peaks which also
propagate from left to right of the plots. The transient peaks,
which are a common feature of these numerical simulations, are a
consequence of the transition from a flat lattice i.e. zero
displacement everywhere at ($t=0$) to one with SAWs ($t>0$).
Moreover, as three membranes are used to generate the SAW, it
takes three periods of the SAW to establish a consistent set of
peaks. In experiments the transients occur too, but many thousands
of SAW wavefronts pass through the split-gate, so the initial
transient peaks are actually insignificant. In the following
numerical experiments, the transients are included in the analysis
as they provide some insight of differing SAW amplitudes, in a
single simulation run.

From Fig.~\ref{fig:potential_X_series}, we see that in this
particular run, the largest amplitude is $\sim 20$~mV and the SAW
resembles the plane wave shape with the expected velocity $2770
\pm 20$~ms$^{-1}$, and a wavelength of $1.0~\mu$m. In our previous
numerical investigations,\cite{Rahman:05} we showed that the
solution method exhibited the characteristic decay into the bulk
as well as a phase difference between the lateral and vertical
displacements $u_x$ and $u_z$,  respectively.\cite{SteveSimon}
From Fig.~\ref{fig:potential_X_series}, we see that despite the
discontinuities in the material parameters due to the the presence
of the heterostructure, we are able to produce coherent waves.
This is consistent with the fact that the electrical and
mechanical properties of Al$_{X}$Ga$_{1-X}$As as shown in Table~I,
differ by less than $1~\%$.

\subsection{\label{sec:resultsb}SAW through a split-gate}

\begin{figure}

\epsfxsize=6.5cm

\centerline{\epsffile{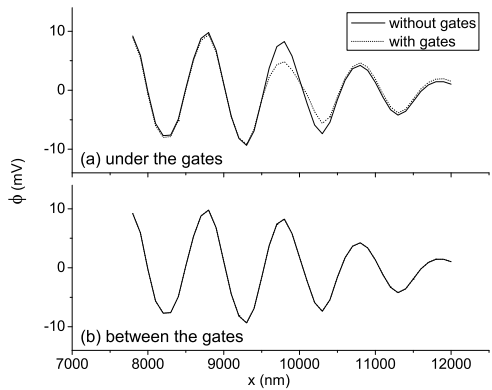}}

\caption{The SAW potential $\phi$ at $t=2.0$~ns, (a) directly
under the gates ($y=500$~nm) (b) in between the gates at the
center of the split-gate constriction ($y=1500$~nm). The curves
were taken $100$~nm below the surface. The SAW amplitude is
reduced directly below the gate as much as $\sim 30~\%$  but less
than $\sim 1~\%$ at the center of the constriction. }
\label{fig:MechEffectOfGates}

\end{figure}

The mechanical effects of the gates are investigated first, by
performing a simulation of a SAW through a split-gate and without
applying any gate voltages. Dirichlet boundary conditions are
therefore not applied to the gates in this case. Moreover, since
here we are interested only in the effect of the purely mechanical
presence of the gates, we can assume that they have dielectric
properties identical to that of the substrate below it. This
allows us to avoid divergences in the solution of
Eq.~(\ref{eqn:PiezoPoisson}) due to absence of the dielectric
parameter $\epsilon^{S}_{ij}$, for the metals.
Figures~\ref{fig:MechEffectOfGates}(a) and
\ref{fig:MechEffectOfGates}(b) compare the SAW potential $\phi$
with and without the mechanical surface gates, directly below the
surface gate and at the center of the constriction, respectively.
Directly below the gate, the SAW amplitude is seen to be reduced
by $\sim 30~\%$, although the amplitude recovers up to at least
$\sim 90~\%$ of its original value after traversing the gate. At
the center of the constriction, the difference of amplitudes
between the gated and un-gated devices is less than one per-cent
in these simulations. The results of
Fig.~\ref{fig:MechEffectOfGates} are consistent with the fact that
electromechanical energy carried by the SAW is much greater than
that could be stored by the gates; the size of the SAW into the
depth is over one micron while the height of the gate is typically
$\sim 0.05~\mu$m.

\begin{figure}

\epsfxsize=8.5cm

\centerline{\epsffile{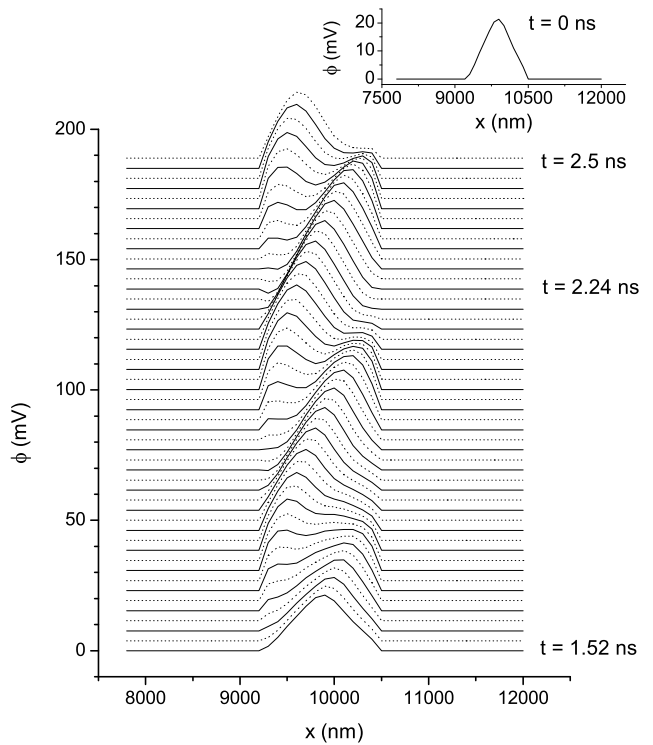}}

\caption{Time development of SAW induced electric potential $\phi$
as a function of the $x$ axis. The simulation includes the
screening by the 2DEG, in the source and drain regions.}
\label{fig:potential_X_series2}

\end{figure}

Having established the mechanical effects of the Ti/Al
split-gates, we now apply an electric potential to the surface
gates such that the barrier height due to the split-gate, after
ignoring static charges on the surface and donor levels, is $\sim
20$~mV at $100$~nm below the surface, corresponding to the low
barrier height and short constriction, regime in experiments. From
the initial time-level of the simulation at $t=0$ to the final at
$t=2.5$~ns, several SAW wavelengths pass through the depleted
region. We observe a pattern from the total SAW and gate potential
$\phi$, resembling that of an `electron
pump',\cite{Pothier:92,Kautz:99} which is repeated with the period
of the SAW. Figure~\ref{fig:potential_X_series2} actually includes
three cycles of this modulation. For the third modulation starting
at $t=2.24$~ns, a local minimum extends between the source 2DEG
and the depleted region. We would expect electrons from the source
2DEG to relax into the minimum. As $t$ increases, a SAW maximum
enters the depleted region forming a potential barrier against
electrons escaping or tunnelling backward into the source 2DEG
while the SAW maximum in front acts against escaping or tunnelling
forward into the drain 2DEG. The electrons are thus confined in
the SAW minimum and are transported along with it, over the
potential barrier formed by the split-gate, until the forward SAW
maximum leaves the depleted region and becomes screened by the
drain 2DEG.

\begin{figure}

\epsfxsize=8.5cm

\centerline{\epsffile{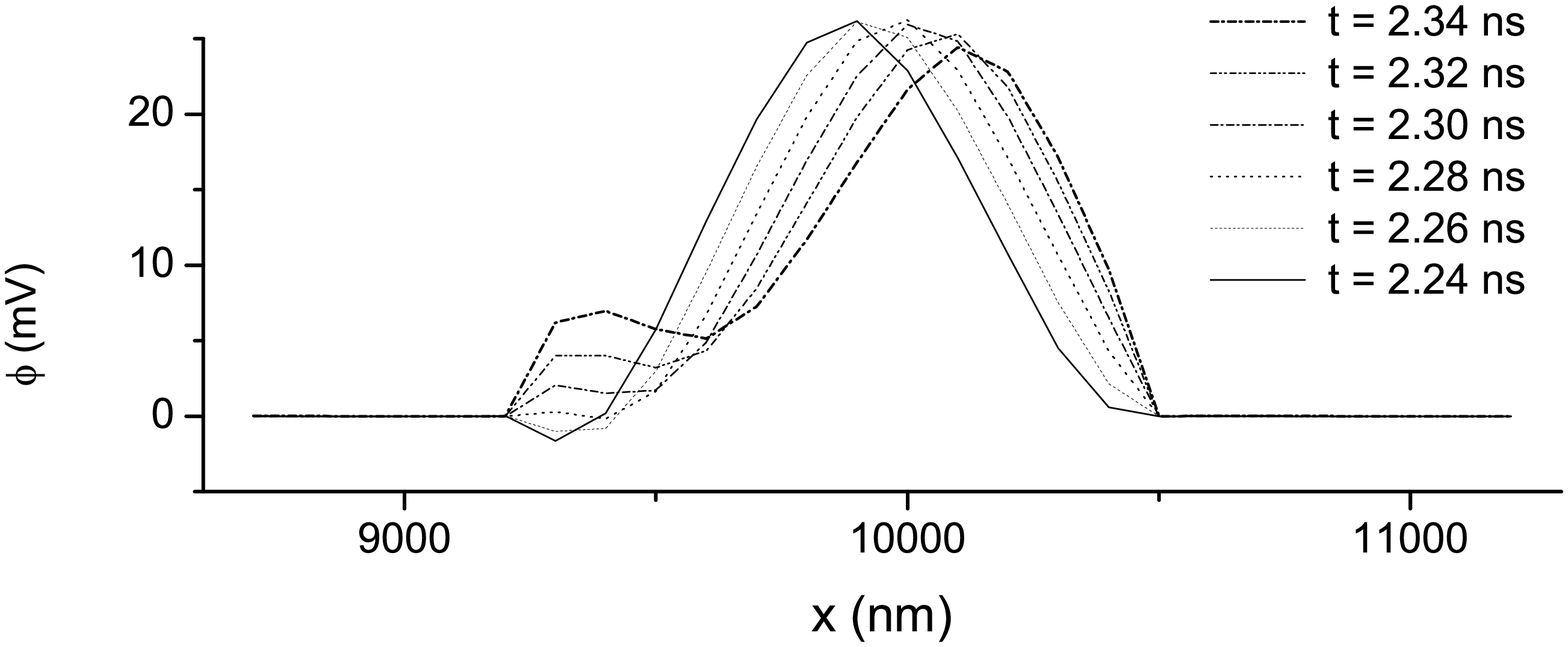}}

\caption{The time development of a rear potential barrier showing
the mechanism by which electrons from the source 2DEG could be
captured.}

\label{fig:SAWcloseUp}
\end{figure}

Figure~\ref{fig:potential_X_series2} suggests that the first two
pump motions, which are caused by the transients in the SAW and
thus have a smaller amplitude, have a lower probability to
transport electrons through the channel, as the potential minima
that extend from the source 2DEG to the depleted region are
smaller, producing overall smaller minima when combined with the
gate potential. Therefore, fewer states will be available in the
dot for the electrons to occupy. For these transients, the rear
tunnel barrier are also smaller and therefore electrons have a
higher probability to escape to the source 2DEG. This is
consistent with Gumbs' investigation,\cite{Gumbs:98} which showed
the acoustoelectric current increase as the \emph{ratio} of the
SAW potential amplitude to the height of the gate induced barrier,
is increased, despite a different reasoning based on analytic
models for the SAW and split-gate potentials.

Experiments have been performed in low SAW power and low barrier
height regime, which utilized SAWs in conjunction which a quantum
dot to effect a pumping modulation of the total electric
potential. \cite{Ebbecke:03}

Figure~\ref{fig:SAWcloseUp} shows the curve plots of the total
electric potential $\phi$ for times $t$ between $2.24$~ns and
$2.34$~ns when a SAW maximum enters the depleted region. It is
clear that the rear potential barrier becomes wider and taller
with time, thus reducing the probability of electrons escaping or
tunnelling backwards into the 2DEG. These results suggest that the
number of electrons transported through the dot is determined
early on in the capture process.

In much of the previous theoretical
work,\cite{Aizin:98,Gumbs:98,Gumbs:99} the increase in width and
height of the rear potential barrier has not formed a significant
part of the analysis. However, Flensberg \emph{et al.}
\cite{Flensberg:99} demonstrated that the rapid change of the SAW
potential barrier at the entrance to the channel results in a
rapid reduction of tunnel-coupling between the source 2DEG and the
SAW minimum inducing non-adiabaticity in the travelling dot.
Flensberg \emph{et al.} then showed that the non-adiabaticity sets
a limit on the accuracy of the quantization plateaus.

In the model of Robinson and Barnes \cite{RobinsonBarnes} the
tunnel barrier decreases with time, and the number of electrons in
the dot is determined when the SAW minima reach the point of
maximum gradient of the split-gate potential barrier. However,
their work was based in the high SAW power and high barrier
height, regime and cosine and gaussian models for the SAW and
split-gate, respectively, were implemented.

\begin{figure}

\epsfxsize=8.5cm

\centerline{\epsffile{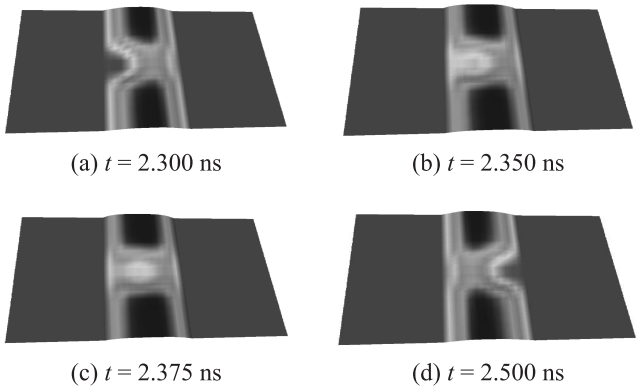}}

\caption{Surface plots showing the time development of the SAW
induced electric potential $\phi$ in the $x$-$y$ plane.}
\label{fig:potential_XY_series}
\end{figure}

Figure~\ref{fig:potential_XY_series} shows surface plots of the
electric potential $\phi$ on a two-dimensional surface on the
$x$-$y$ plane, $100$~nm below the surface at a sequence of times
$t$ providing a more vivid illustration of the pumping mechanism.
In particular, at $t=2.35$~ns a quantum well begins to form on the
left hand side of the split-gate as a SAW minimum enters the
constriction. At $t=2.375$~ns the SAW minimum is located in
between the gates, forming a well defined quantum well of a
circular geometry (although the geometry will vary from circular
to elliptical depending on the split-gate potential). At
$t=2.5$~ns, the SAW minimum exits the constriction, and the well
begins to disappear.

\begin{figure}

\epsfxsize=7.25cm

\centerline{\epsffile{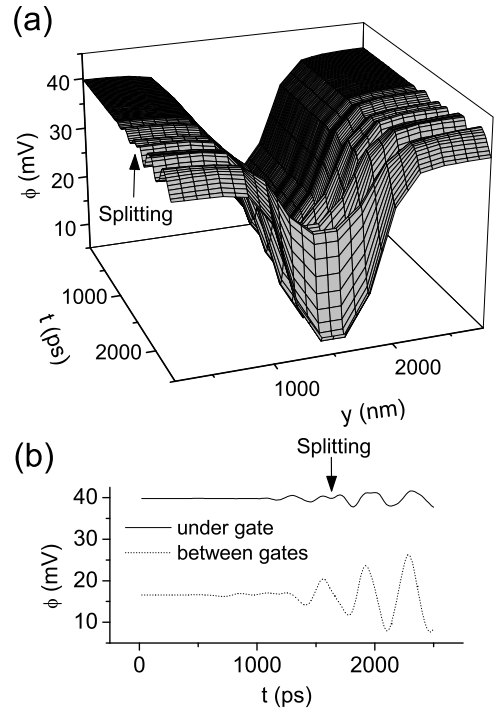}}

\caption{(a) Time development of the total electric potential
$\phi$ as a function of the $y$ coordinate. (b) The electric
potential $\phi$ directly under the gate and in between the gate.}

\label{fig:potential_Y_series2}

\end{figure}

Figures~\ref{fig:potential_Y_series2}(a) and
\ref{fig:potential_Y_series2}(b) show curve-plots of the total
electric potential $\phi$ parallel to the $y$-axis, through the
center of the constriction. From
Fig.~\ref{fig:potential_Y_series2}(a), one can see the
oscillations of the total electric potential $\phi$ as the SAW
propagates through it. The amplitude of the oscillations at the
center of the constriction is $\sim 20$~mV, about the same
amplitude as that of the bare SAW (i.e., without gates), whereas
the SAW amplitude below the gate is $\sim 5$~mV and therefore has
been reduced by $\sim 75~\%$. This is due to the screening of the
SAW potential by the gates, the presence of mechanical gates as
discussed earlier, and to a much lesser extent, due to the
mechanical strains caused by the gate voltage affecting the SAW
motion.\cite{Rahman:05}$^{,}$ \cite{ft1} These figures also show
the a change in the SAW frequency under the gates, in the form of
`splitting' of the peaks, where labelled. In particular, the SAW
oscillates at a higher frequency under gates.

The peak splitting or increase in the frequency of the SAW under
the gates is likely to be caused by the difference in elasticity
constants of the gates and substrate. In particular, the
elasticity constants of the gates are greater that of the
substrate. The gates are therefore much `stiffer' than the
substrate, and have a higher natural frequency of vibration. The
vibration of surface gates impart additional vibrational
components to the underlying SAW. A more detailed understanding of
the mechanism behind the peak splitting would involve
investigation of the lateral and vertical components of the
displacements under a gate (and perhaps experiments with different
gate materials), and therefore a major digression from the scope
and intention of this paper but a possible topic of future work.

The simulated changes in amplitude and frequency of SAW would be
undesirable for electron transport experiments as they may allow
both tunnel and Coulomb interactions between electrons in
consecutive SAW minima. However, these phenomena seem to be more
significant for the first three peaks passing through the gate,
than for the last peak. The last peak has a larger amplitude and
thus greater energy, and is therefore more resistant to these
effects.

Finally, from Fig.~\ref{fig:potential_Y_series2}, one can see the
absence of any significant turbulence in the SAW motion along the
$y$ axis, which may have affected systems of one-dimensional
channels in parallel which are currently under
development.\cite{Rodriguez:05,Kataoka:05} This also suggests that
diffraction of the SAW on propagation though our split-gate is
beyond the accuracy of our numerical work i.e. extremely small.
Therefore the only waves in our system are plane waves which
travel parallel to the lateral edges of the computational domain
and are not reflected of them.

\section{\label{sec:conclusions}CONCLUSION} \label{sec:Conclusion}

We have performed numerical simulations in order to investigate
the dynamics resulting from the propagation of a piezoelectric SAW
though a Q1DC defined by metallic split-gates on the surface of a
GaAs/Al$_{X}$Ga$_{1-X}$As heterostructure. Our simulations were
performed in three spatial dimensions fully incorporating the
mechanical and electrical parameters of the materials, and were
based on the strongly screened, low SAW power and short
constriction regime. We ignored the presence of static charges, as
we are interested in the dynamical properties of the SAW as it
propagates through a realistic split-gate device.

We have demonstrated significant `amplitude reduction' of the SAW
potential, up to $30$~\% due to the mechanical presence of the
surface gates, as well as up to $75$~\% screening of the SAW
potential, by the biased split-gate device. In addition, the
simulations demonstrated the recovery of the SAW, after amplitude
reduction due to the presence of mechanical gates, and also the
absence of significant damping or screening effects at the center
of the constriction formed by the split-gate. These effects have
been demonstrated theoretically for the first time for a realistic
device, and would be difficult, if possible at all, to achieve
without a numerical procedure such as ours. The results suggest
that the coherent propagation of SAWs through systems of Q1DCs in
series or in parallel could also be achieved, although for
definiteness, further simulations may be required.

Through the incorporation of a simple model for the screening by
the 2DEG, we demonstrated a total electric potential modulation
that resembled an electron pump and provided a simple model for
the capture of electrons by the SAW, from the source 2DEG, and the
release of the electrons to the drain 2DEG.

\section{\label{sec:acknowledgements}ACKNOWLEDGEMENTS}

We thank G. Gumbs, V. I. Talyanskii, A. M. Robinson, H. K.
Bhadeshia, C. G. Smith, J. Jefferson and C. J. B Ford for comments
and useful discussions. This work was partly funded by the UK
EPSRC and QIP IRC. SR and MK acknowledge the Cambridge-MIT
Institute for financial support.


\begin{references}
\bibitem{ft2} More realistic boundary conditions
at the interface (between the substrate and any existing vacuum)
may consist of either the continuity of Dz or charges = 0, but the
nature of the results presented here are such that they would not
be affected by such changes to the boundary conditions.
\bibitem{rotation}  To achieve this, we rotate the crystal by $45$ degrees around
the $z$ axis.
\bibitem{ft1} The strains produced by the voltages applied to the gates in
this lower barrier regime, are small compared to those produced by
the moving SAW and hence are not shown.
\end{references}

\end{document}